\documentclass[aps,prb,superscriptaddress,preprint]{revtex4}  
\usepackage{amsmath} 
\usepackage{graphicx} 
\usepackage{makeidx} 

\begin{document}

\title{Quantum circuits based on coded qubits encoded in chirality of electron spin complexes in triple quantum dots} 

\author{Chang-Yu Hsieh} 
\affiliation{Quantum Theory Group, 
Institute for Microstructural Sciences, 
National Research Council, Ottawa, Canada K1A 0R6} 
\affiliation{Department of Physics, 
University of Ottawa, Ottawa, ON, Canada, K1N 6N5} 

\author{Pawel Hawrylak} 
\affiliation{Quantum Theory Group, 
Institute for Microstructural Sciences, 
National Research Council, Ottawa, Canada K1A 0R6} 
\affiliation{Department of Physics, 
University of Ottawa, Ottawa, ON, Canada, K1N 6N5}

\begin{abstract} 

We present  a theory of quantum circuits based on logical qubits encoded in chirality of electron spin complexes in  lateral gated semiconductor triple quantum dot molecules with one electron spin in each dot.  Using microscopic Hamiltonian we show how  to initialize, coherently control and measure  the quantum state of a chirality based coded qubit using static in-plane magnetic field and  voltage tuning of individual dots. The microscopic model of two interacting coded qubits is established and mapped to   an Ising Hamiltonian,  resulting in  conditional two-qubit phase gate.  
\end{abstract}

\maketitle


\section{Introduction} \label{sec:INTRO}
There is currently interest in exploiting electron spin for nano-spintronic\cite{sachrajda_hawrylak_book2003} and quantum information processing\cite{brum_hawrylak_sm1997,korkusinski_hawrylak_book_2008,loss_divincenzo_pra1998, hanson_kouwenhoven_2007}. This is partly motivated by electron spin long coherence times \cite{hanson_witkamp_prl2003} and availability of scalable semiconductor technology.  In the simplest approach, a physical qubit is identified with the two states of an electron spin, which can be manipulated by applying local magnetic fields. Much progress has been achieved using micro-magnet\cite{pioro-ladriere_tokura_apl2007,pioro_obata_nat2008} technology with electron spin qubits. An alternative approach is to encode a logical qubit in a two level system constructed with spin complexes\cite{korkusinski_hawrylak_book_2008}. This includes singlet-triplet two electron qubit\cite{levy_prl2002} and a logical qubit encoded in a degenerate ground state of a three spin complex\cite{divincenzo_bacon_nature2000}. As suggested by DiVincenzo et al.\cite{divincenzo_bacon_nature2000}, in the framework of a Heisenberg model for a logical qubit encoded in a linear chain of three spins, such a logical qubit can be manipulated by the control of the exchange interaction between pairs of spins. The ability to manipulate spin state with a voltage is related to the relation between the orbital and spin part of the many-electron wavefunction.  A detailed microscopic model of a three electron complex in a single semiconductor quantum dot (QD), exploring the orbital and spin relation, has been investigated 
and compared with experiment by one of us\cite{hawrylak_prl1993}. A similar microscopic model of three electrons localized in a lateral gated triangular triple quantum dot molecule in the plane of a GaAs/GaAlAs heterojunction was proposed and explored by one of us in Ref.~\onlinecite{hawrylak_korkusinski_ssc2005,gimenez_korkusinski_prb2007}. The coded qubit was identified with
chirality of three electron spin complex, or equivalently, two possible directions of a minority spin motion.  
Other proposals to implement coded qubits with both triple quantum dots\cite{weinstein_hellberg_pra2005_2} (TQDs) and atom traps\cite{georgeot_mila_arxiv2009, trif_troiani_prl2008}also exist. 

The advantages of working with a TQD-based coded qubit are two-fold.  First, every quantum gate can be implemented electrically.  In such a scheme, magnetic field will only be used for initialization and (or) measurement of a coded qubit.  Second, the coded qubit involves Decoherence Free Subspace\cite{bacon_kempe_prl2000,viola_fortunato_sci2001,weinstein_hellberg_pra2005_2,lidar_whaley_book2008} (DFS) and is immune to channels of collective decoherence. This reduces decoherence of a TQD-based coded qubit due to charge fluctuations as discussed in Ref.~\onlinecite{gimenez_hsieh_prb2009}. 

Other interesting phenomena involving TQD molecules include non-Fermi liquid behavior when coupled to leads\cite{delgado_hawrylak_prb2008,ingersent_ludwig_prl2005}, potential for generating maximally entangled three-partite GHZ and W states\cite{eibl_kiesel_prl2004,ghosh_kar_njp2002}and manipulation of total spin\cite{shim_delgado_prb2009}. 

In this work we develop a microscopic theory of quantum circuits based on coded qubits encoded in chirality of electron spin complexes in TQD. We use a combination of Linear Combination of Harmonic Orbitals - Configuration Interactions (LCHO-CI)\cite{gimenez_korkusinski_prb2007}, Hubbard and Heisenberg models to determine a set of optimal conditions for single qubit operations and describe the two qubit gate. We show that there exists an in-plane magnetic field direction and magnitude optimal for single qubit operations. However, the magnetic field will rotate qubit states in an undesirable way; and we show how this rotation can be controlled by tuning voltages on the gates. Once the optimal sets of magnetic fields and voltages have been obtained, the exact diagonalization techniques are used to establish and verify an effective two-qubit Hamiltonian. It is shown that the effective two qubit interaction is Ising-like, leading to a two-qubit phase gate. The present work establishes both single qubit and two-qubit operations necessary for performing quantum computation using TQD-based coded qubits.

The paper is organized as follows.   Sec. \ref{sec:INTRO} contains introduction. Sec. \ref{sec:SYS} describes the quantum circuit based on chirality of 3 electron spin complexes and computational methodology.  In Sec. \ref{sec:CQBT}, we present a definition of a TQD-based coded qubit, discuss initialization, single qubit operations using voltages, and measurement of the state of the coded qubit.  In Sec. \ref{sec:DQOP}, we present an effective Hamiltonian of 2 coupled coded qubits and show that it can be canonically transformed to an Ising interaction.  Sec. \ref{sec:CON} contains summary.

\section{The Model} \label{sec:SYS}
\begin{figure}
\centering
\includegraphics[width=0.9\textwidth]{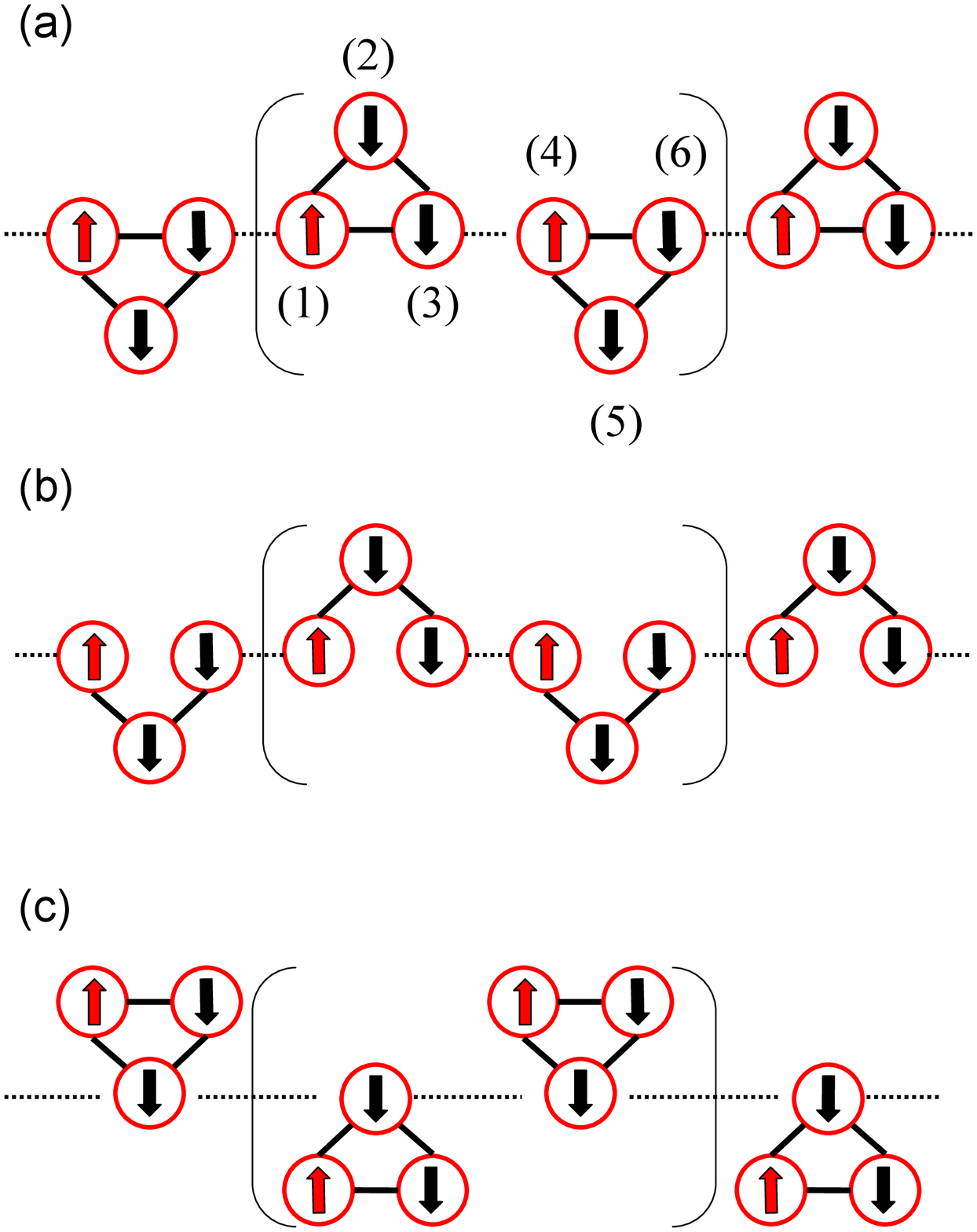}
\caption{ Three possible layouts for a chain of TQD-based coded qubits.  Each solid and dash line represents an exchange interaction between a pair of QDs.  The layouts (a) and (b) differ in that the constituent
coded qubits are implemented with triangular TQD in (a) and with linear TQD in (b).  The layouts (a) and (c) differ in the way logical qubits are connected in a chain.  If the inter-qubit distance is reduced beyond a threshold, an additional set of exchange interactions will arise and dominate the behavior of the system in layout (c). }
\label{fig:layout}
\end{figure}

Fig.(\ref{fig:layout}a) shows schematically layout of quantum circuits based on coded qubits encoded in chirality of electron spin complexes in triangular TQDs. The circles denote individual lateral quantum dots formed in the 2D electron gas (2DEG) at the heterojunctions of AlGaAs/GaAs by metallic gates on the AlGaAs surface. The gates are set to confine a single electron in each dot denoted by an arrow. Additional gates (not shown here) are used to control tunneling between dots in the same TQD molecule, shown schematically as solid lines. Dashed lines indicate tunneling between neighboring TQDs.  The tunneling between any pair of QDs is responsible for exchange interaction of electron spins localized on each QD. The brackets indicate two TQDs isolated from the rest of the circuit. It is assumed that 
any number of TQDs can be isolated from the rest of the circuit by turning off the boundary exchange interactions with gate voltages.

For comparison, Fig.(\ref{fig:layout}b) shows the circuit composed of linear, instead of triangular, TQDs, similar to the linear chain of spins  proposed by DiVincenzo et al., in which the coded qubits in the chain are implemented with only two exchange interactions.  

Fig.(\ref{fig:layout}c) is another possible architecture studied by Weinstein\cite{weinstein_hellberg_pra2005_2} et. al. for spin system. However, in this design, bringing TQDs close together induces interactions between quantum dots beyond the ones indicated by dashed line.

In this work we focus mainly on triangular TQDs-based quantum circuit shown in Fig.(\ref{fig:layout}a).
Since electrons are well localized in each QD, a system of two TQDs in a chain can be very well described by an extended Hubbard model\cite{korkusinski_gimenez_prb2007}. With  $c^{\dag}_{i\sigma} (c_{i\sigma}) $ electron creation (annihilation) operator for the  electron with spin $\sigma=\pm1$ on the $i-th$ QD, the Hubbard Hamiltonian reads:
\begin{eqnarray}
\label{eq:NHubbard}
\hat{H}_{2} &=& \sum_{n=1}^{2}\sum_{i=3n-2}^{3n}\sum_\sigma E_{i\sigma} \hat{n}_{i\sigma} + \sum_{n=1}^{2} \sum_{\substack{i, j = 3n-2\\ i \neq j }}^{3n} \sum_ \sigma t_{ij} \hat{c}^{\dag}_{i\sigma}\hat{c}_{j\sigma} \nonumber \\
 &  & + \frac{1}{2}\sum_{n=1}^{2} \sum_{\substack{i, j = 3n-2\\ i \neq j }}^{3n}V \hat{\rho}_{i} \hat{\rho}_{j} + \sum_{n=1}^{2} \sum_{i=3n-2}^{3n}U \hat{n}_{i\uparrow}\hat{n}_{j\downarrow}  \nonumber \\ 
&  & + \sum_{\sigma} t' \left(\hat{c}^{\dag}_{3\sigma}\hat{c}_{4\sigma}+\hat{c}^{\dag}_{4\sigma}\hat{c}_{3\sigma}\right)
+ V' \hat{\rho}_{3} \hat{\rho}_{4},
\end{eqnarray} 
where $n=1,2$ labels the two TQD molecules from left to right, indices $i,j$ range from $3n-2$ to $3n$ label each QD from left to right in the $n-th$ TQD molecule.  The intra-TQD Hubbard parameters $t_{ij}$, $V$, $U$ and $E_{i\sigma}$ are the tunneling matrix element between the $i-th$ and the $j-th$ QDs, the off-site Coulomb interaction parameter between any two QDs in the same TQD molecule, the on-site Coulomb interaction strength for any QD, and  the on-site energy for the $i-th$ QD respectively. The on-site energy depends on spin and magnetic field which is applied in the plane of a TQD molecule.
The Hubbard parameters $t'$ and $V'$ represent the inter-molecular tunneling matrix element and inter-molecular Coulomb interactions.  $\hat{n}_{i\sigma}=\hat{c}^{\dag}_{i\sigma}\hat{c}_{i\sigma} $ is the number operator for the $\sigma$ electron on the $i-th$ QD.  $\hat{\rho}_i = \hat{n}_{i\uparrow} + \hat{n}_{i\downarrow}$ is the electron charge density operator on the $i-th$ QD.  When we discuss triangular resonant TQDs, we drop the subscripts on all the parameters. In our model, we consider the parameters corresponding to regime of strong correlations: $t'_{ij} < t_{ij}  << V' < V < U$. 

The intra-TQD Hubbard parameters, $E_i$, $ t_{ij}$, $V$, and  $U$,  in Eq.(\ref{eq:NHubbard}) are obtained from a microscopic calculation for single TQD based on LCHO-CI method as explained in Ref.~\onlinecite{shim_hawrylak_prb2008}. The inter-TQD Hubbard parameter $V'$ is taken to be the direct Coulomb interaction between 2 charges localized on adjacent edge dots of two neighboring TQD molecules. 

The energy spectrum and eigenstates of the Hubbard Hamiltonian for one and two TQD molecules are obtained using Configuration Interaction technique. For a given number of electons $N_e$ we construct all possible
configurations $\vert k={i_{N_e}\sigma_{N_e},..,i_2\sigma_2,i_1\sigma_1}\rangle= \hat{c}^{\dag}_{i_{N_e} \sigma_{N_e}}... \hat{c}^{\dag}_{i_2\sigma_2}\hat{c}^{\dag}_{i_1\sigma_1}\vert 0 \rangle$, build Hamiltonian matrix in the space of configurations, and diagonalize it numerically.

At half-filling, the low-energy spectrum of the Hubbard model can be approximated by a spectrum of a Heisenberg model\cite{korkusinski_gimenez_prb2007,scarola_park_prl2004, scarola_dassarma_pra2005} describing electron spins localized in each dot:
\begin{equation}
\label{eq:Heis1}
\hat{H}_{heis} =  \sum_{ i < j } J_{ij}  \mathbf{S}_i \cdot \mathbf{S}_j.
\end{equation}
The exchange interactions, $J_{ij}$, for the TQDs can be expressed in terms of tunneling matrix elements and 
quantum dot energies:

\begin{equation}
\label{eq:SingleExchange}
J_{ij} = 2\vert t_{ij} \vert ^2\left(\frac{1}{U-V+(E_i-E_j)}+\frac{1}{U-V-(E_i-E_j)}\right).
\end{equation}  
The exchange interaction can be controlled by either tuning the tunneling matrix element $t_{ij}$ by, for example, additional gates controlling the height of the tunneling barrier, or by biasing 
the dots and changing their on-site energy $E_{i(j)}$.

\section{Coded qubit encoded in chirality of electron spin complex in a triple quantum dot molecule} \label{sec:CQBT}
\begin{figure}
\centering
\includegraphics[width=0.9\textwidth]{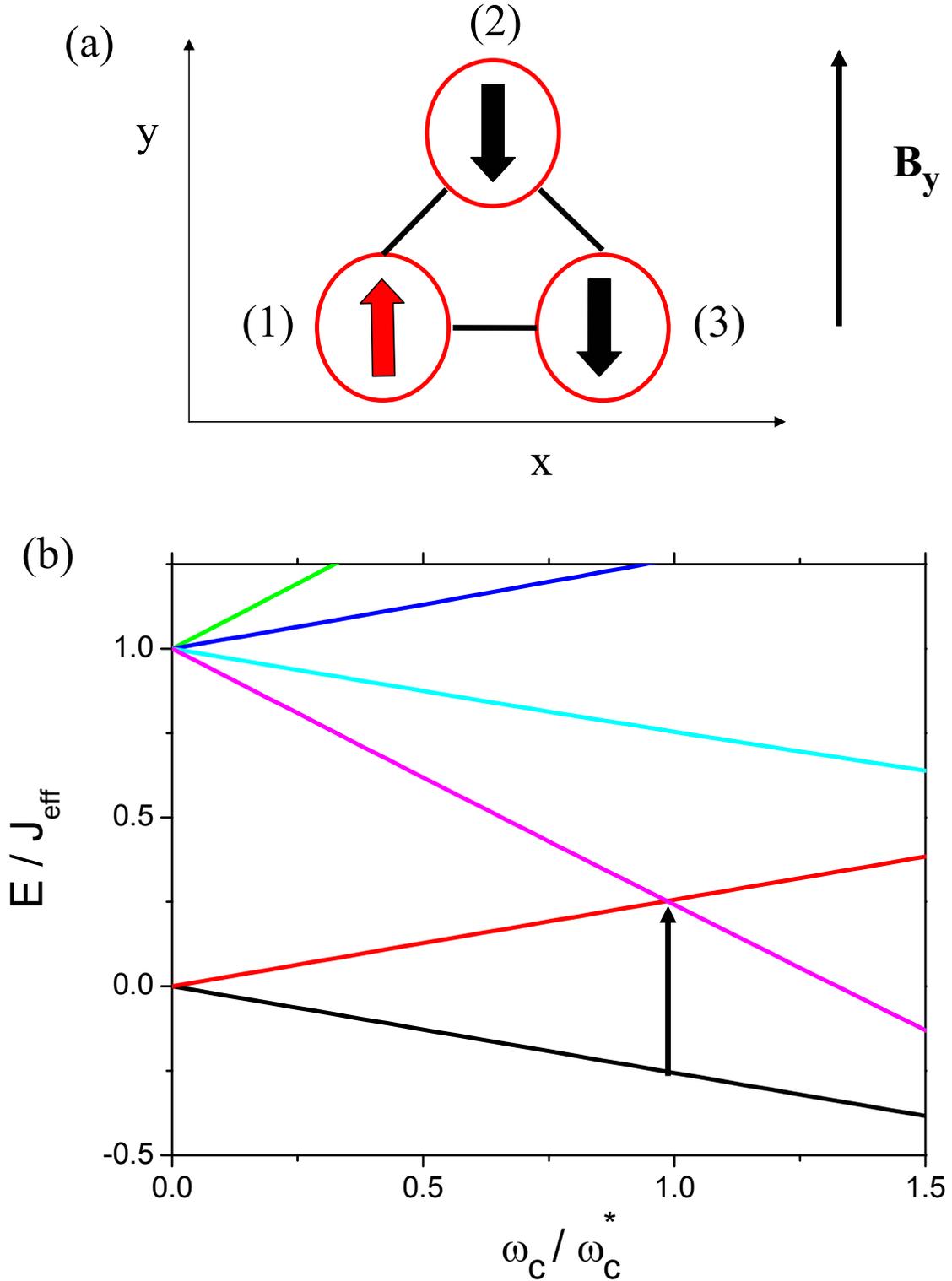}
\caption{(a). A schematic of a single TQD initialized in $S_y = -1/2$ subspace under a $B_y$ field.  The $B_y$ field breaks the discrete rotational symmetry, and it affects QD 2 differently from QDs 1 and 3. (b). The energy spectrum of a triangular TQD as a function of $\omega_c / \omega_c^*$, where $\omega_c$ is the cyclotron frequency used in LCHO-CI calculation.  $\omega_c^*$ corresponds to a magnetic field $B^*$ such that the gap between computational subspace (the lowest energy level in the plot) and the rest of spectrum is maximized as indicated by the black arrow.  }
\label{fig:bfield}
\end{figure}

In this section, we discuss a single coded qubit shown in Fig.(\ref{fig:bfield}a): its preparation, initialization, operation and measurement in the presence of a lateral magnetic field in the $y$ direction.

The qubit is encoded in quantum states of a three electron spin complex in a fully symmetric and half-filled TQD.
An example of one of the three possible configurations $\left| \uparrow \downarrow \downarrow \right\rangle= 
\hat{c}^{\dag}_{1 \uparrow}\hat{c}^{\dag}_{2 \downarrow} \hat{c}^{\dag}_{3\downarrow}|0>$ with $S_y=-1/2$ is shown schematically in Fig.(\ref{fig:bfield}a). The three configurations with the minority spin on QD 1, 2 or 3 form a doubly degenerate ground state with total spin $S=1/2$ and an excited state with total spin $S=3/2$
separated from the ground state by $J_{eff}$, as shown in Fig.(\ref{fig:bfield}b).

The doubly degenerate ground state with fixed $S_y=-1/2$ forms an effective two level system. We identify the coded qubit levels $\vert q_{+} \rangle$ and $\vert q_{-} \rangle$ with
\begin{eqnarray}
\label{eq:ChiralStates}
\left| q_s \right\rangle & = & \frac{1}{\sqrt{3}}\sum_{n=1}^{3} e^{i 2\pi  n s / 3} \left| n \right\rangle,
\end{eqnarray}
where $s=\pm$ and  $\left| n \right\rangle = S_n^+ \left| \downarrow \downarrow  \downarrow  \right\rangle$.
The three spin complex is characterized by chirality  $\chi =  \left( \mathbf{S}_1 \times \mathbf{S}_2 \right) \cdot \mathbf{S}_3$ which measures the degree of collinearity of the three spins.  The two coded qubit levels, Eq.(\ref{eq:ChiralStates}), are the eigenstates of chirality operator with eigenvalues $\chi= \pm \sqrt{3}/4 $.  
These two states also can be more intuitively characterized by minority spin moving either to the left or to the right as in Resonant Valence Bond (RVB) plaquette 
\cite{wen_wilczek_prb1989, fradkin_book, tatara_garcia_prl2003, bulaevskii_batista_prb2008}. 

We note that in the absence of the magnetic field the coded qubit states in $S_y=1/2$ subspace are degenerate with coded qubit states in $S_y=-1/2$ subspace. If we are to work in the computational space corresponding to $S_y=-1/2$, any process which flips the electron spin will remove the coded qubit from its computational space.
In order to separate the computational Hilbert space $S_y=-1/2$ from $S_y=+1/2$ subspace, a magnetic field $B_y$ is applied along the $y$ direction as shown in Fig(\ref{fig:bfield}a). We avoid applying magnetic field applied in the $z$ direction, because it activates undesirable, higher order spin-spin interactions\cite{scarola_park_prl2004, scarola_dassarma_pra2005}, whereas the magnetic field applied in the plane only modifies the on-site energies of QDs: for  $B_y$ field,  QDs 1 and 3 energy level as $E_{1(3)}(B)= E_{1(3)}(0)+1/2 \omega_c (R/2)^2+ g\mu B S_y$, where $R$ is the spatial separation of QDs 1 and 3 and $\omega_c = eB_y/m^*_ec$ is the cyclotron frequency, while the QD 2 energy level only acquires the Zeeman term. Hence magnetic field in the $y$ direction effectively lowers the energy of the QD 2 by $\Delta E_2= - 1/2 \omega_c (R/2)^2$ with respect to QDs 1 and 3.

Fig.(\ref{fig:bfield}b) shows the energy spectrum of the coded qubit as a function of $\omega_c $ obtained in LCHO-CI method.  In our LCHO-CI calculation,  we use the following parameters for a specific resonant TQD.  The inter-dot distance is 10 $a_B^*$, where  $a_B^*=9.79 nm$ is the effective Bohr radius for $GaAs$.  The confining Gaussian potential on the $i-th$ QD is of the form, $V_i(x,y) = -V_i^0 \exp\left(-((x-x_i)^2+(y-y_i)^2)/d_i^2\right)$, where $V_i^0=4.0 Ry^*$ and $d_i=2.5 a_B^*$. $ Ry^*=5.93 meV$ is the effective Rydberg for $GaAs$.  In Fig.(\ref{fig:bfield}b), the optimal cyclotron frequency $\omega_c^*$ corresponds to $B^* = 0.69 T$.

\begin{figure}
\centering
\includegraphics[width=0.7\textwidth]{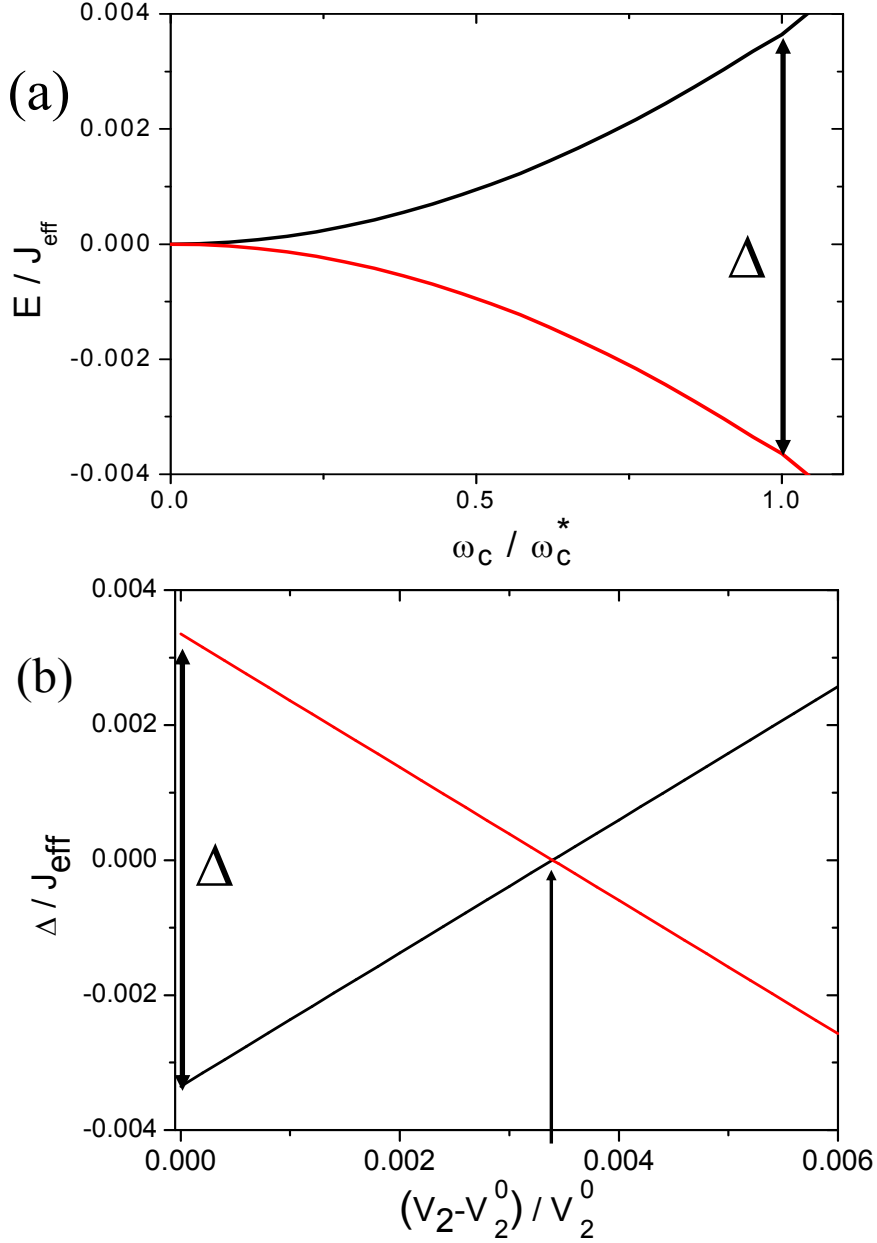}
\caption{ (a). The energy spectrum of qubit levels (the zoom in of the lowest level in Fig. (2b)) as a function of $\omega_c / \omega_c^*$.  The $B_y$ field splits the doubly degenerate ground state, and $\Delta$ corresponds to the gap between the two levels at $\omega_c^*$ or  equivalently at $B_y^*$.  (b). The energy spectrum of the qubit levels under $B_y^*$ as a function of  voltage $V_2$ on QD2, and $V_2^0$ is the unbiased voltage at resonant condition when there is no external $B$ field.  The splitting of levels under $B$ field can be restored via gate voltage tuning.}
\label{fig:gate}
\end{figure}

At $B=0$, the four-fold degenerate ground state is separated from the four-fold degenerate spin polarized excited state by $J_{eff}$. The two computational Hilbert spaces corresponding to
$S_y = \pm 1/2$ separate energetically with increasing magnetic field while the spin polarized states decrease in energy. At magnetic field $B^*$, such that $ J_{eff}= 2 g \mu B^*$, the energy gap is maximized
and should correspond to the working point that can maintain the longest coherence time of the coded qubit. The coded qubit should operate at this value of magnetic field. However, as discussed above, the magnetic field effectively biases the QD 2.
This removes the degeneracy of the two qubit levels and rotates them from their zero magnetic field states.
The energy splitting of the two coded qubit levels as a function of applied magnetic field is shown in Fig.(\ref{fig:gate}a).
The splitting is a fraction of the large energy scale $J_{eff}$. For the largest gap, i.e. $B=B^*$, the splitting is $\Delta$. In order to restore the degeneracy of the two coded qubit levels, one can apply voltage to the QD 2. 
In Fig.(\ref{fig:gate}b),  microscopic calculations done in LCHO-CI show that a positive voltage bias on the QD 2 can indeed bring the TQD back on resonance in the presence of $B$ field.
We have now established the coded qubit and the best conditions for its operation.

In order to operate the coded qubit, we need to be able to initialize it.
We propose to initialize the coded qubit by turning off both interaction between QDs 1 and 3 and interaction between QDs 1 and 2. The only remaining interaction is between QDs 2 and 3.
The ground state of a TQD,
$\left| L_0  \right\rangle  = \vert \downarrow_1 \rangle  \vert S_{23} \rangle  $,
becomes  a product of a spin down state of an electron on the QD 1 and a singlet state of electrons across QDs 2 and 3. This intuitively is a ground state in magnetic field. The singlet state of two electrons in a pair of QDs 2 and 3 can be generated in real time starting from two electrons in a biased QD 3.
This procedure does not generate directly the coded qubit levels $\vert q_ {+} \rangle$ and $\vert q_{-} \rangle$, but the state  $\left| L_0  \right\rangle $ is a linear superposition of the two qubit levels:
$\left| L_0  \right\rangle  = {-i \over{\sqrt{2}}} ( e^{i2\pi/3} \vert q_- \rangle -e^{-i2\pi/3}\vert q_+ \rangle ) $. Once interactions are turned on, any state can be obtained from the initial state.
This can be seen by writing the TQD Heisenberg Hamiltonian in the basis of the two coded qubit levels $\left\{ \vert q_{\pm} \rangle \right\}$ ,  
\begin{equation}
\label{eq:singleCodedQubit}
\hat{H}_{1q} = \frac{1}{2}\left(J_{12} - \frac{1}{2}J_{13}- \frac{1}{2}J_{23}\right) \hat{\sigma}_x + \frac{\sqrt{3}}{4}\left( J_{13}-J_{23}\right) \hat{\sigma}_y.
\end{equation}
If we take $J_{13}=J_{23}$, and let $2J_{12} > J_{13}+J_{23}$, then the Heisenberg Hamiltonian corresponds to $\hat{\sigma}_x$ operation.  If we take $J_{23} > J_{13}$ and $2J_{12} = J_{13} + J_{23}$, then the coded qubit Hamiltonian Eq.(\ref{eq:singleCodedQubit}) corresponds to $\hat{\sigma}_y$ operation.  The capability to rotate a qubit with respect to two different axes on a Bloch sphere allowes us to generate arbitrary single qubit operations.

In practice, we tune the exchange interaction $J_{ij}$ through biasing QDs and changing their energies.  As long as the biasing $|\Delta E| << U-V$ is satisfied, we expect the quantum state to remain in the qubit subspace during the process of tuning the exchange interactions as discussed in Ref.~\onlinecite{weinstein_hellberg_pra2005_2}.

Next, we discuss the measurement of coded qubits.  Several proposals\cite{bulaevskii_batista_prb2008, cao_hu_pla2008, georgeot_mila_arxiv2009} discusses methods of detecting chirality of a triangular (three-body) antiferromagnetic cluster.  For our specific TQD-based coded qubit, we propose to apply a $B_z$ field to split the two coded qubit levels with different chirality.  As discussed earlier, the application of $B_z$ field will give rise to additional terms proportional to chirality operator on top of the Heisenberg model, Eq.(\ref{eq:Heis1}), used for qubit modeling.  Since the additional term to the Heisenberg model is strictly proportional to the chirality, it splits the two
coded qubit levels with the gap given by
\begin{equation}
\label{eq:chiralgap}
\frac{12\sqrt{3}|t|^3}{(U-V)^2}\sin\left( \frac{2\pi\Phi}{\Phi_0}\right),
\end{equation}
where $\Phi$ is the magnetic flux through the device and $\Phi_0 = eh/c$ is the flux quanta.  This energy gap can be interpreted as the energy difference between two magnetic dipole moments\cite{bulaevskii_batista_prb2008} oriented in opposite directions under $B_z$. In Ref.~\onlinecite{scarola_dassarma_pra2005}, Scarola et. al. also showed that the unintended qubit rotation due to the additional 
chirality term can be significantly reduced for a range of specific weak $B_z$ fields that satisfy $R/l_h \geq 1$ and $\omega_c/\omega_o  \leq 1$, where $l_h = \sqrt{\hbar c/eB}(1+4\omega_o^2/\omega_c^2)^{-1/4}$ is the modified magnetic length, and $\omega_o$ is the confining frequency used to approximate the potential of a QD.  Thus, initializing the system under $B_z$ might not be a good idea as the field has to stay on for the entire period of quantum computation and eventually modify the quantum state of the coded qubit due to magnetic moment coupling with $B_z$.  However, a measurement done with advanced spectroscopy only requires $B_z$ to be turned on for a relatively short time and significantly limit the extent to which the coded qubit will be modified.

\section{Double qubit operations} \label{sec:DQOP}
We now turn to discuss double coded qubit operations using both Hubbard and Heisenberg models.
Starting with the Hubbard Hamiltonian, Eq.(\ref{eq:NHubbard}), we derive perturbatively a Heisenberg Hamiltonian 
for a complex of six electron spins with a set of exchange interactions $J_{ij}$
\begin{eqnarray}
\label{eq:DoubleExchange}
J_{12} = J_{56} = J_a   & = & \frac{4t^2}{U-V} ,\nonumber \\
J_{13} = J_{23} = J_{45} = J_{46} = J_b & = &  \frac{2t^2}{(U-V+V')}+\frac{2t^2}{U-V-V'} , \nonumber \\
J_{34} = J_c  & = & \frac{4t'^2}{U-V'},
\end{eqnarray}
with all other exchange interactions set to zero. For all the Hubbard parameters considered in this study (i.e. weak inter-TQD tunneling and strong Coulomb interaction), $J_c$ is usually about two order of magnitudes smaller than $J_a$ and $J_b$, mainly due to the fact that $t' <  t / 2$ in our study.

We treat the inter-TQD exchange interaction, scaled by $J_c$, perturbatively to derive an effective interaction in the coupled coded qubits subspace, $\left\{ \vert q_+^L\rangle \vert q_+^R\rangle,\vert q_+^L\rangle \vert q_-^R\rangle, \vert q_-^L\rangle \vert q_+^R\rangle,\vert q_-^L\rangle \vert q_-^R\rangle \right\}$, where the superscript $L$ and $R$ stand for left and right TQD respectively. 
The effective qubit-qubit interacting Hamiltonian reads
\begin{equation}
\label{eq:codedqubit4}
\hat{H}_{2q} = 4|\alpha'|(\hat{n} \cdot \vec{S}_1)(\hat{n} \cdot \vec{S}_2)+2|\beta'|\hat{m} \cdot \vec{S}_1+2|\beta'|\hat{m} \cdot \vec{S}_2,
\end{equation}
where
$ \alpha' =  e^{-\frac{i2\pi}{3}}J_c/9,$ and 
$ \beta'   =  e^{-\frac{i\pi}{3}}\left(-J_a/2+J_b/2+J_c/27\right),$
and the direction vectors in Eq.(\ref{eq:codedqubit4}) are defined as follows:
$ \hat{n} =  \left(\sqrt{|\alpha'|+Re(\alpha')}/\sqrt{2|\alpha'|}, \sqrt{|\alpha'|-Re(\alpha')|}\sqrt{2|\alpha'|},0\right),
\hat{m} = \left( Re(\beta')/|\beta'|, Im(\beta')/|\beta'|, 0 \right)$.
Thus, the effective interaction for 2 TQD-based code qubits  is equivalent to an $XY$ Hamiltonian under a uniform in-plane magnetic field for 2 spins.

If we rotate the effective qubit-qubit interacting Hamiltonian, Eq.(\ref{eq:codedqubit4}), from the coded qubit basis to the Jacobian basis, $\left\{ \vert L_0^L\rangle \vert L_0^R\rangle,\vert L_0^L\rangle \vert L_1^R\rangle, \vert L_1^L\rangle \vert L_0^R\rangle,\vert L_1^L\rangle \vert L_1^R\rangle \right\}$,  the rotated Hamiltonian reads
\begin{equation}
\label{eq:2qubitL}
\hat{H}_{2Lq} = \alpha S_1^zS_2^z + \beta S_1^z + \beta S_2^z , 
\end{equation}
where $\alpha =J_c / 9 $, and $\beta = -J_a/2 + J_b/2 + J_c/18$ .  The Jacobian state $\vert L_0 \rangle$ is defined in Sec.\ref{sec:CQBT} when we discuss the initialization of the coded qubit.  The Jacobian state $\vert L_1\rangle = \sqrt{1/3}\vert \downarrow_1 \rangle \vert T_{23}^0 \rangle - \sqrt{2/3} \vert \uparrow_1 \rangle \vert T_{23}^- \rangle$, where $\vert T_{23}^0\rangle$ is a $S_z=0$ triplet state of electrons on QDs 2 and 3 and   $\vert T_{23}^-\rangle$ is a $S_z=-1$ triplet state of electrons on QDs 2 and 3.  Similarly, $\vert L_1 \rangle =  -1/\sqrt{2} ( e^{i2\pi/3} \vert q_- \rangle + e^{-i2\pi/3}\vert q_+ \rangle )$ can also be written as a linear combination of the coded qubit levels.  Therefore this rotation of basis only requires single qubit operations applied to each coded qubit; the two Hamiltonians, Eq.(\ref{eq:codedqubit4}) and Eq.(\ref{eq:2qubitL}), are locally equivalent\cite{makhlin_qip2002}.  As Ising interaction can be used to generate CNOT gate\cite{nielsen_chuang_book}, we show how to generate non-local two qubit interactions with 2 TQD-based coded qubits.

\begin{figure}
\centering
\includegraphics[width=0.9\textwidth]{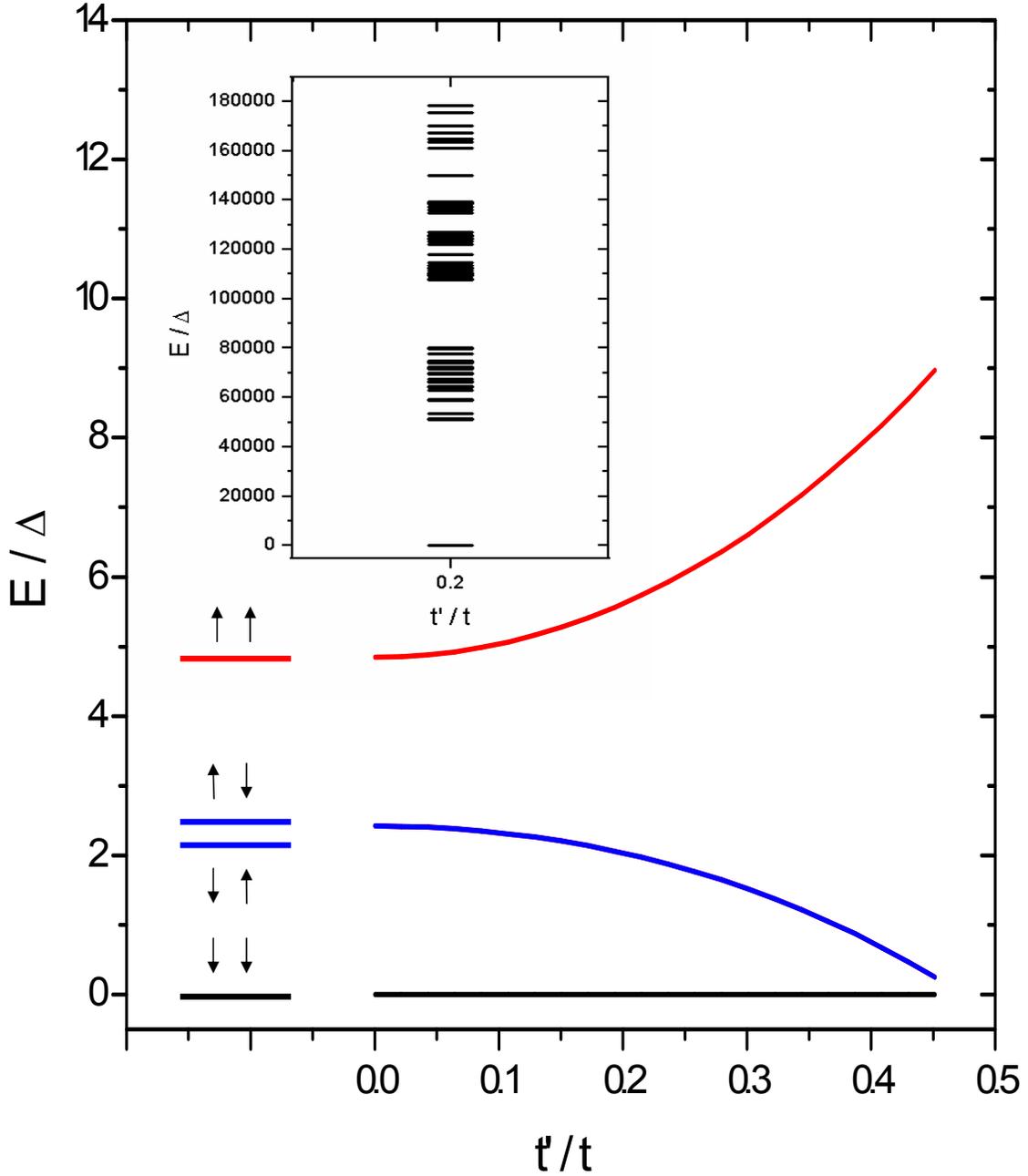}
\caption{The four lowest energy levels of 2 coupled triangular TQDs as a function of inter-TQD tunneling $t'$.  The energy spectrum resembles that of the Ising model with an external field:  a doubly degenerate levels corresponding to state $\vert \downarrow \uparrow \rangle$ and $\vert  \uparrow \downarrow \rangle$ and two unique levels $\vert \downarrow \downarrow \rangle$ and $\vert \uparrow \uparrow \rangle$. "$X2$" denotes a doubly degenerate level.
Inset: The entire energy spectrum at $t' / t = 0.2$ calculated with LCHO-CI method.}  
\label{fig:ising}
\end{figure}

Fig.(\ref{fig:ising})  shows results of exact diagonalization of the Hubbard Hamiltonian for two coupled TQDs as a function of increasing coupling $t'$ between the two molecules. The inset shows the entire energy spectrum at $t'/t=0.2$. The low energy spectrum consists of four levels, characterized mostly by the two lowest levels of each TQD.
These  4 low-lying energy levels  are well separated from the rest of the spectrum and constitute the two coupled coded qubit subspace.  The low-lying spectrum contains a doubly
degenerate level, which is a signature of the Ising model in an external field.  In practice, the tuning of $J_c$, the inter-TQD exchange interaction, should be done via tuning the tunneling parameter $t'$.   Fig. (\ref{fig:ising}) shows that the Ising model features of the energy spectrum are maintained over a wide range of values of $t'$, and hence the Ising model Hamiltonian, derived above, describes the coupling of two coded qubits very well.

\section{Conclusion} \label{sec:CON}
In summary, we present a theory of quantum circuits based on coded qubits encoded in chirality of electron spin complexes in  lateral gated semiconductor triple quantum dot molecules with one electron spin in each dot.  Using microscopic Hamiltonian and exact diagonalization techniques we show how to initialize, coherently control and measure the quantum state of a chirality based coded qubit using static in-plane magnetic field and  voltage tuning of individual QDs. The microscopic model of two interacting coded qubits is established and mapped to an Ising Hamiltonian. Hence both conditional two-qubit phase gate and  voltage controlled single qubit operations are demonstrated.

\section{Acknowledgment} \label{sec:Ack}
The authors thank NSERC, QUANTUMWORKS, CIFAR, NRC-CNRC CRP and OGS for support.
C.-Y. Hsieh would like to thank Y.-P. Shim and A. Sharma for useful discussions.



\newpage

\end{document}